\documentclass{emulateapj}

\newcommand{\eg}{e.g.,\ }

\def\eg{{\it e.g.,\ }}
\def\simgt{\ {\raise-.5ex\hbox{$\buildrel>\over\sim$}}\ }
\def\simlt{\ {\raise-.5ex\hbox{$\buildrel<\over\sim$}}\ }

\def \h70{{h_{70}}}

\slugcomment{For submission to Astrophysical Journal Letters.}
\shorttitle{SFR vs M$_*$ Scatter Evolution}
\shortauthors{Kurczynski et al.}

\usepackage{graphicx,subfigure}

\DeclareGraphicsExtensions{.pdf,.png,.jpg,.eps}

\usepackage{amssymb, amsmath}

\usepackage[hyperindex,breaklinks]{hyperref}

\begin{document}


\title{Evolution of Intrinsic Scatter in the SFR-Stellar Mass Correlation at 0.5$<$z$<$3}


\author{Peter Kurczynski\altaffilmark{1}, Eric Gawiser\altaffilmark{1}, Viviana Acquaviva\altaffilmark{2}, Eric F.~Bell\altaffilmark{3}, Avishai Dekel\altaffilmark{4}, Duilia F.~de Mello\altaffilmark{5,6}, Henry C.~Ferguson\altaffilmark{7}, Jonathan P.~Gardner\altaffilmark{5}, Norman A. Grogin\altaffilmark{7}, Yicheng Guo\altaffilmark{9}, Philip F. Hopkins\altaffilmark{15,16}, Anton M. Koekemoer\altaffilmark{7}, David C. Koo\altaffilmark{9}, Seong-Kook Lee\altaffilmark{10}, Bahram Mobasher\altaffilmark{11}, Joel R.~Primack\altaffilmark{12}, Marc Rafelski\altaffilmark{5,13}, Emmaris Soto\altaffilmark{5,6}, Harry I. Teplitz\altaffilmark{14}}


\altaffiltext{1}{Department of Physics and Astronomy, Rutgers University, Piscataway, NJ 08854, USA}
\altaffiltext{2}{New York City College of Technology, Brooklyn, NY 11201, USA} 
\altaffiltext{3}{Department of Astronomy, University of Michigan, Ann Arbor MI 48109, USA} 
\altaffiltext{4}{Center for Astrophysics and Planetary Science, Racah Institute of Physics, The Hebrew University, Jerusalem 91904, Israel}
\altaffiltext{5}{Laboratory for Observational Cosmology, Astrophysics Science Division, Code 665, Goddard Space Flight Center, Greenbelt, MD 20771, USA}
\altaffiltext{6}{Department of Physics, The Catholic University of America, Washington, DC 20064, USA}
\altaffiltext{7}{Space Telescope Science Institute, 3700 San Martin Drive, Baltimore, MD 21218, USA}
\altaffiltext{9}{UCO/Lick Observatory, Department of Astronomy and Astrophysics, University of California, Santa Cruz, CA, USA}
\altaffiltext{10}{Center for the Exploration of the Origin of the Universe, Department of Physics and Astronomy, Seoul National University, Seoul, 151-742, Republic of Korea}
\altaffiltext{11}{Department of Physics and Astronomy, University of California, Riverside, CA, USA}
\altaffiltext{12}{Department of Physics, University of California, Santa Cruz, CA, 95064, USA}
\altaffiltext{13}{NASA Postdoctoral Program Fellow}
\altaffiltext{14}{Infrared Processing and Analysis Center, MS 100-22, Caltech, Pasadena, CA 91125.}
\altaffiltext{15}{TAPIR, Mailcode 350-17, California Institute of Technology, Pasadena, CA 91125, USA}
\altaffiltext{16}{Department of Astronomy and Theoretical Astrophysics Center, University of California Berkeley, Berkeley, CA 94720}


\begin{abstract}

We present estimates of intrinsic scatter in the Star Formation Rate (SFR) - Stellar Mass (M$_*$) correlation in the redshift range $0.5 < z < 3.0$ and in the mass range $10^7 < M_* < 10^{11}$ M$_\odot$.  We utilize photometry in the {\it Hubble} Ultradeep Field (HUDF12) and Ultraviolet Ultra Deep Field (UVUDF) campaigns and CANDELS/GOODS-S.  We estimate SFR, M$_*$ from broadband Spectral Energy Distributions (SEDs) and the best available redshifts.  The maximum depth of the HUDF photometry  (F160W  29.9 AB, 5$\sigma$ depth) probes the SFR-M$_*$ correlation down to M$_* \sim$10$^7$ M$_\odot$, a factor of 10-100$\times$ lower in M$_*$ than previous studies, and comparable to dwarf galaxies in the local universe.  We find the slope of the SFR-M$_*$ relationship to be near unity at all redshifts and the normalization to decrease with cosmic time.  We find a moderate increase in intrinsic scatter with cosmic time from 0.2 to 0.4 dex across the epoch of peak cosmic star formation.  None of our redshift bins show a statistically significant increase in intrinsic scatter at low mass.  However, it remains possible that intrinsic scatter increases at low mass on timescales shorter than $\sim$100 Myr.  Our results are consistent with a picture of gradual and self-similar assembly of galaxies across more than three orders of magnitude in stellar mass from as low as 10$^7$ M$_\odot$. 
\end{abstract}


\section{INTRODUCTION}
\label{IntroductionSection}

A central issue in understanding how galaxies form is whether star formation is a gradual, continuous process or whether it happens in bursts.  The widely reported correlation between Star Formation Rate (SFR) and stellar mass (M$_*$) in star-forming galaxies (``main sequence"; \eg \citealt{2007ApJ...660L..43N,2007ApJS..173..267S,2007ApJ...670..156D,2011ApJ...738..106W}) provides an observational means to address this issue.  Because M$_*$ is related to past-average SFR, the small total observed scatter around this correlation ($\sim$ 0.3 dex at z $\lesssim$ 2; \citealt{2013ApJ...770...57B}) suggests gradual assembly of stellar mass, as opposed to bursty star formation. 

Bursty star formation introduces scatter to the SFR-M$_*$ relation and diversity to star formation histories (\eg \citealt{2014ApJ...785L..36A}); it is found to dominate the evolution of low-mass galaxies in simulations \citep{2015MNRAS.451..839D,2014ApJ...792...99S} and in observations of local galaxies (\eg \citealt{2015MNRAS.450.4207B,2014MNRAS.441.2717K,2014ApJ...789..147W}).  In particular, \citet{2010ApJ...721..297M} find starbursts in dwarf galaxies  to occur with durations in the 100 Myr - 1 Gyr range that will be probed here.  Furthermore, stochasticity in star formation may arise at low SFR values due to sampling effects \citep{2012ApJ...745..145D,2014MNRAS.444.3275D,2011ApJ...741L..26F}.

Guides to the extensive SFR-M$_*$ literature can be found in \citet{2013ApJ...770...57B} and \citet{2014ApJS..214...15S}.  Studies to date have not modeled scatter.  The typically reported {\it total observed scatter} includes SFR and M$_*$ measurement uncertainties, and covariances as well as the underlying intrinsic scatter.\footnote{\citet{2015ApJ...799..183S} and \citet{2015ApJ...804..149S} do compute intrinsic scatter post hoc from fit residuals, without covariances or estimated uncertainties to the intrinsic scatter.}  However, cosmological galaxy evolution simulations make predictions for the physically meaningful quantity, intrinsic scatter, which in the absence of measurement errors and covariances, is the standard deviation (dex) of the SFR-M$_*$ fit residuals.   In this Letter, we present an analysis of the SFR-M$_*$ relation that specifically addresses intrinsic scatter.  

To probe SFR-M$_*$ to the lowest possible mass, we utilize photometry from the {\it Hubble Space Telescope (HST)} in the Hubble Ultradeep Field (HUDF;  \citealt{2006AJ....132.1729B}), including HUDF12 (\citealt{2013ApJ...763L...7E,2013ApJS..209....3K}; see also \citealt{2013ApJS..209....6I}), UVUDF \citep{2013AJ....146..159T} and the Cosmic Assembly Near-infrared Deep Extragalactic Legacy Survey (CANDELS; \citealt{2011ApJS..197...35G,2011ApJS..197...36K}).  Magnitudes are in the AB system; we use the cosmology $\Omega_\Lambda$ = 0.7, $\Omega_0$ = 0.3, and H$_{0}$ = 70 km s$^{-1}$ Mpc$^{-1}$.

%
%
\section{DATA AND SAMPLE SELECTION}
\label{DataSection}

We form samples of galaxies for analysis from  CANDELS GOODS-S and UVUDF photometric catalogs.  We utilize the selection criteria of \citet{2015ApJ...801...97S}
to reject poor quality data, stars and AGN.  We utilize exceptionally deep HUDF photometry since our primary motivation is to probe to low mass, while the larger, complimentary CANDELS data provides overlapping and continuous coverage of the mass range up to $\sim10^{11}$ M$_\odot$.  Notably, the {\it HUDF} photometry enables detecting dwarf galaxies with M$_* \sim 10^{7}$ M$_\odot$ at $z>0.5$ (compare with the Small Magellanic Cloud, M$_* \sim 10^{8}$ M$_\odot$).  

%
%
We select sources in the redshift range $0.5 < z \leq 3.0$, including 2444 spectroscopic redshifts.  In the larger CANDELS catalog, we require spectroscopic redshifts; cross-listings in the smaller UVUDF catalog \citep{2015AJ....150...31R} use grism (3D-HST; \citealt{2014ApJS..214...24S}) or photometric redshifts. UVUDF photometric redshifts have fewer outliers than CANDELS photometric redshifts \citet{2015AJ....150...31R}.  Using the best-available redshifts is preferable for estimating scatter, see Section \ref{ResultsSection}.  We form samples in five redshift bins $0.5 < z \leq 1.0$, $1.0 < z \leq1.5$, $1.5 < z \leq 2.0$, $2.0 < z \leq 2.5$, $2.5 < z \leq 3.0$ that have 1369, 1100, 673, 439, 435 sources respectively.

%
We use 17 bands from the CANDELS photometry (U-band through IRAC; \citealt{2013ApJS..207...24G}) 
to generate input data for fits to the Spectral Energy Distributions (SEDs), which are used to estimate physical parameters such as SFR and M$_*$.  

%
\section{METHOD}
\label{MethodSection}

We fit SEDs with a Markov Chain Monte Carlo based program \citep{2011ApJ...737...47A,2012IAUS..284...42A} to estimate SFR, M$_*$, $E(B-V)$ (spectral reddening), age and star formation history timescale, $\tau$ (discussed below).  Age varies from 1 Myr to the age of the universe at each (binned) galaxy redshift.  Ages are found to fall between $\sim$100 Myr - 1 Gyr, with no ages younger than 10 Myr; however, the 1 Myr lower limit was found to improve $\chi^2$ for several sources compared to a more stringent 10 Myr lower limit.

Galaxy mass varies between 10$^{4}$$-$10$^{15}$M$_\odot$; $E(B-V)$ varies from $0.01-0.99$.  $\tau$ is sampled logarithmically from $0.02-4.99$ Gyr.  We use \citet{2003MNRAS.344.1000B} stellar templates, including nebular emission lines, \citet{1959ApJ...129..608S} Initial Mass Function (IMF), and \citet{2000ApJ...533..682C} dust attenuation law.  Metallicity is fixed at $Z=0.2 Z_\odot$; fits at solar metallicity have generally poor convergence and larger parameter uncertainties.  We utilize parameter uncertainties and covariances for each galaxy.  

We explore several continuous  star formation histories including constant, linear, exponential (``$\tau$ model'') and linear-exponential (``delayed $\tau$ model'').  The linear-exponential model \citep{2010ApJ...725.1644L} permits both rising and falling star formation, and yields comparable median $\chi^2$ values as the next best model (exponential).  We report results obtained with this model, and estimate its parameters, $t_0$ (time-to-peak) and $\tau$ (decay timescale).  Instantaneous SFRs  are most sensitive to star formation within $\sim100's$ Myr before observation.  These SFRs are less sensitive to short timescale (e.g.~10 Myr) variations than spectroscopic indicators (H$\alpha$), and yield lower scatter than them \citep{2014MNRAS.445..581H,2015MNRAS.451..839D}.

%
%
%
%
%
%
%
%
%
%

We reject SEDs with bad fits ($\chi^2 > 50$; 263, 229, 186, 158, 80 galaxies in each redshift bin, respectively), or poor convergence (GR $> 0.2$; \citealt{Gelman1992}; 131, 95, 55, 62, 30 galaxies).  SED fits with large $\chi^2$ values have potentially under-estimated parameter uncertainties which overestimates scatter.\footnote{a small effect; rejecting $\chi^2 > 100$ increases scatter by $\sim10$\%}   Our final redshift-binned samples have 958, 692, 466, 246, and 326 galaxies for SFR-M$_*$ analysis.  

For each sample, we fit log SFR and log M$_*$ values to the model:
\begin{equation}
\textnormal{log}~SFR = a \times \textnormal{log}~M_* + b + N(0,\sigma_{IS})
\end{equation}
The parameters $a$ and $b$ describe the linear relationship and the Gaussian random variable, $N(0,\sigma_{IS})$, with zero mean and unknown standard deviation, $\sigma_{IS}$, describes intrinsic scatter.   We use the analytic method of \citet{Fuller1987}, F87 hereafter, to estimate parameters in the presence of uncertainties and covariances.  A full-width tenth maximum clipping range is obtained from the histogram of initial fit residuals to exclude outliers (our results are insensitive to the details of clipping).  We re-fit the outlier-clipped data to estimate the model parameters.

\section{RESULTS}
\label{ResultsSection}

\begin{figure}[h]
\begin{center}
\subfigure[$0.5<z\leq1.0$]{%
	\includegraphics[scale=0.40]{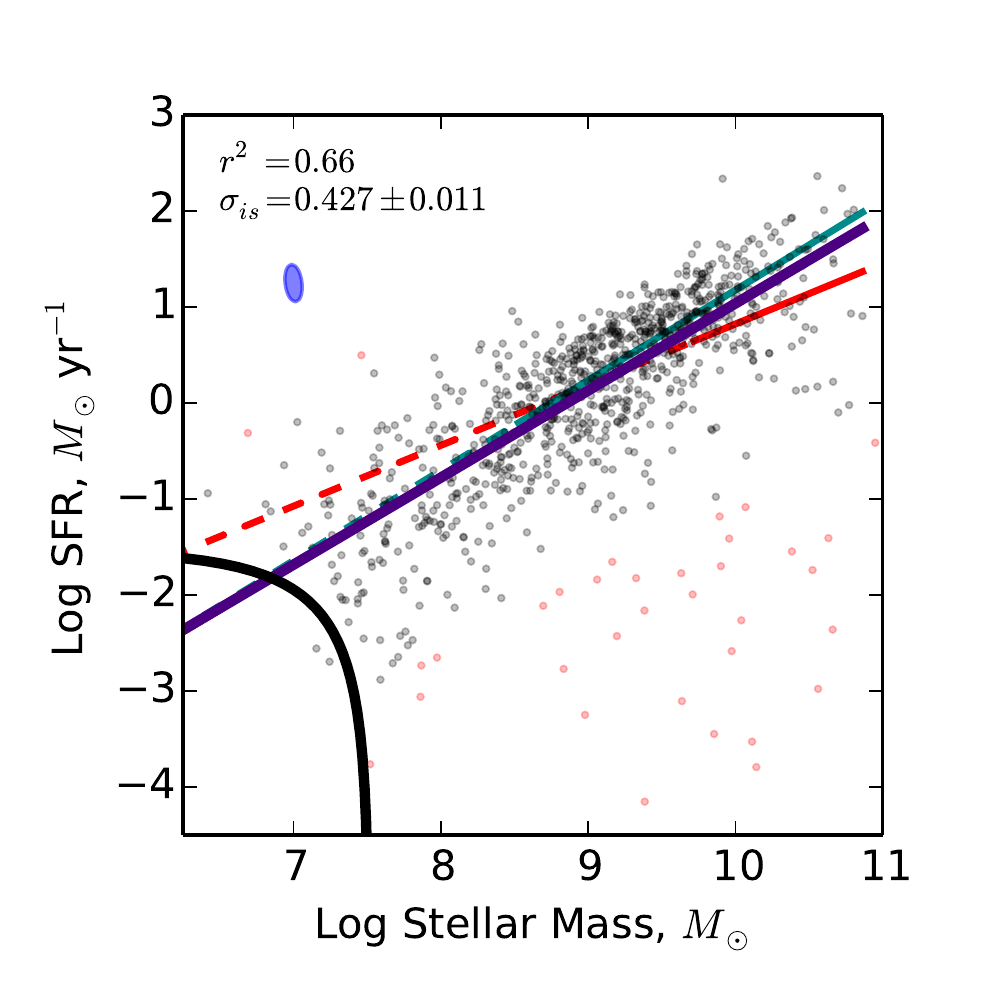}
}
\subfigure[$1.0<z\leq1.5$]{%
	\includegraphics[scale=0.40]{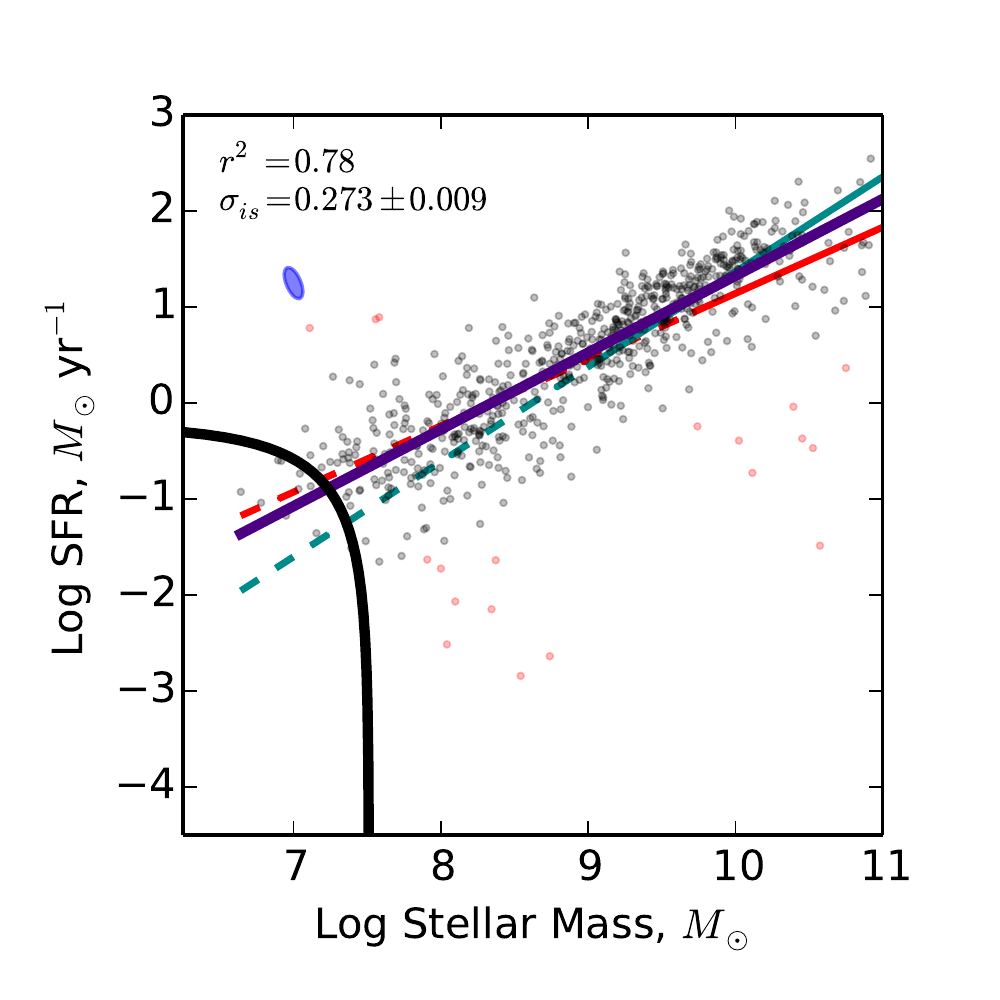}
}
\subfigure[$1.5<z\leq2.0$]{%
	\includegraphics[scale=0.40]{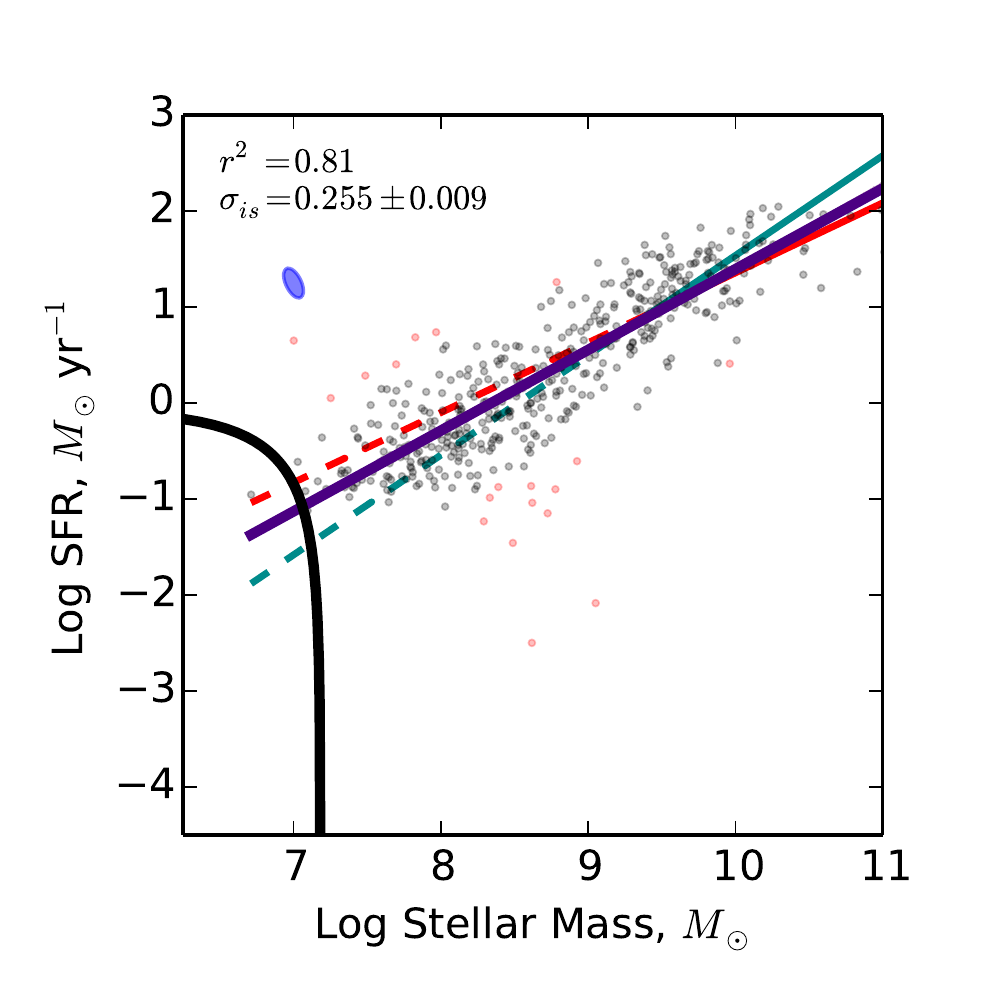}
}
\subfigure[$2.0<z\leq2.5$]{%
	\includegraphics[scale=0.40]{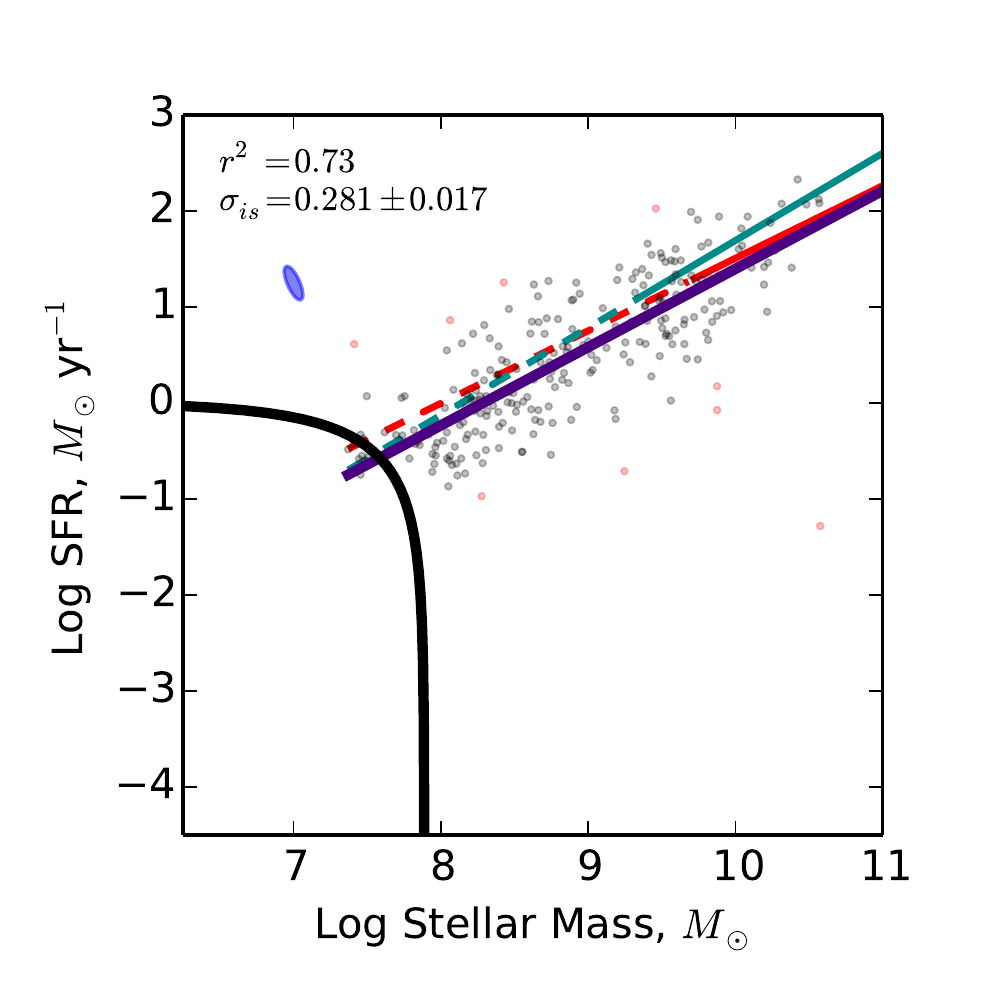}
}
\subfigure[$2.5<z\leq3.0$]{%
	\includegraphics[scale=0.40]{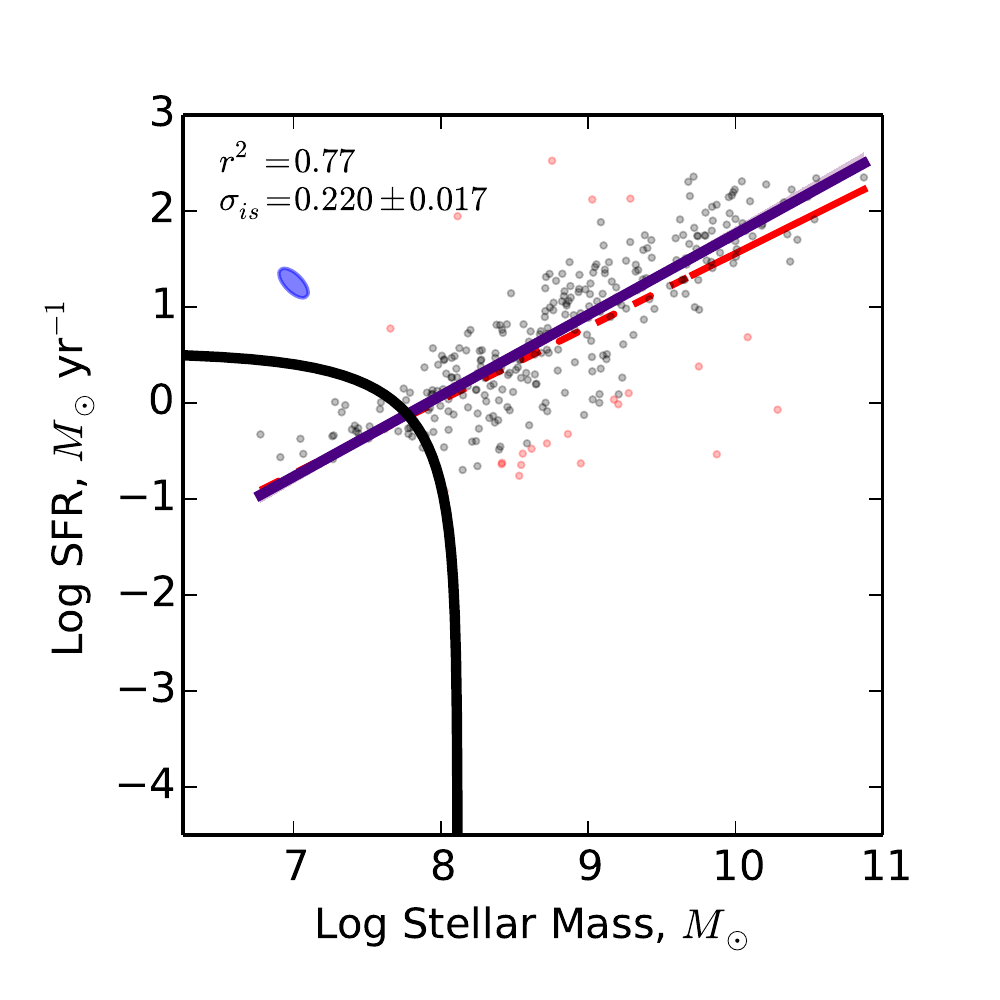}
}

\end{center}
\caption{Star Formation Rate vs. Stellar Mass (M$_*$) in the redshift range $0.5 < z < 3.0$ for galaxies in combined CANDELS (spec-z) and UVUDF (photo-z) sample.  In the $0.5 < z \leq 1.0$ bin, seven outliers with log SFR $< -3$ are not shown.  Outliers (red points) from an initial fit are clipped; remaining galaxies (gray points) are used to determine the best fit (dark purple).  Results from Whitaker et al.~(2014; cyan) and the meta-analysis of Speagle et al.~(2014; red) are shown; dashed regions indicate extrapolations from their reported ranges in M$_*$.  Selection curves are shown in black; our data are insensitive to galaxies that would fall to the lower left of each curve.  The squared Pearson correlation coefficient, $r^2$, and estimated intrinsic scatter, $\sigma_{IS}$, (dex) are indicated by the text label.  A typical error ellipse is shown in the upper left, with half-width and half-height equal to the median error in log M$_*$ and log SFR respectively, and orientation determined by the median covariance.}  
\label{FIGURE:01}
\end{figure}

Results include estimated parameters for five redshift-binned samples spanning the mass range $10^7 \lesssim M_* \lesssim 10^{11}$ M$_\odot$.  Figure \ref{FIGURE:01} shows the SFR vs M$_*$ data and fits; we find significant correlations (Pearson $r^2$ values in the range $0.66 - 0.81$).  We compare with \citet{2014ApJ...795..104W}\footnote{adjusted upwards by a factor of $\textnormal{log}_{10}(1.8)$ to convert from their adopted \citet{2003PASP..115..763C} IMF to the \citet{1959ApJ...129..608S} IMF used here.} over their redshift range $0.5 < z < 2.5$, and the meta-analysis of \citet{2014ApJS..214...15S}.  

\begin{figure}[h]
\begin{center}
\subfigure[$0.5<z\leq1.0$]{%
	\includegraphics[scale=0.40]{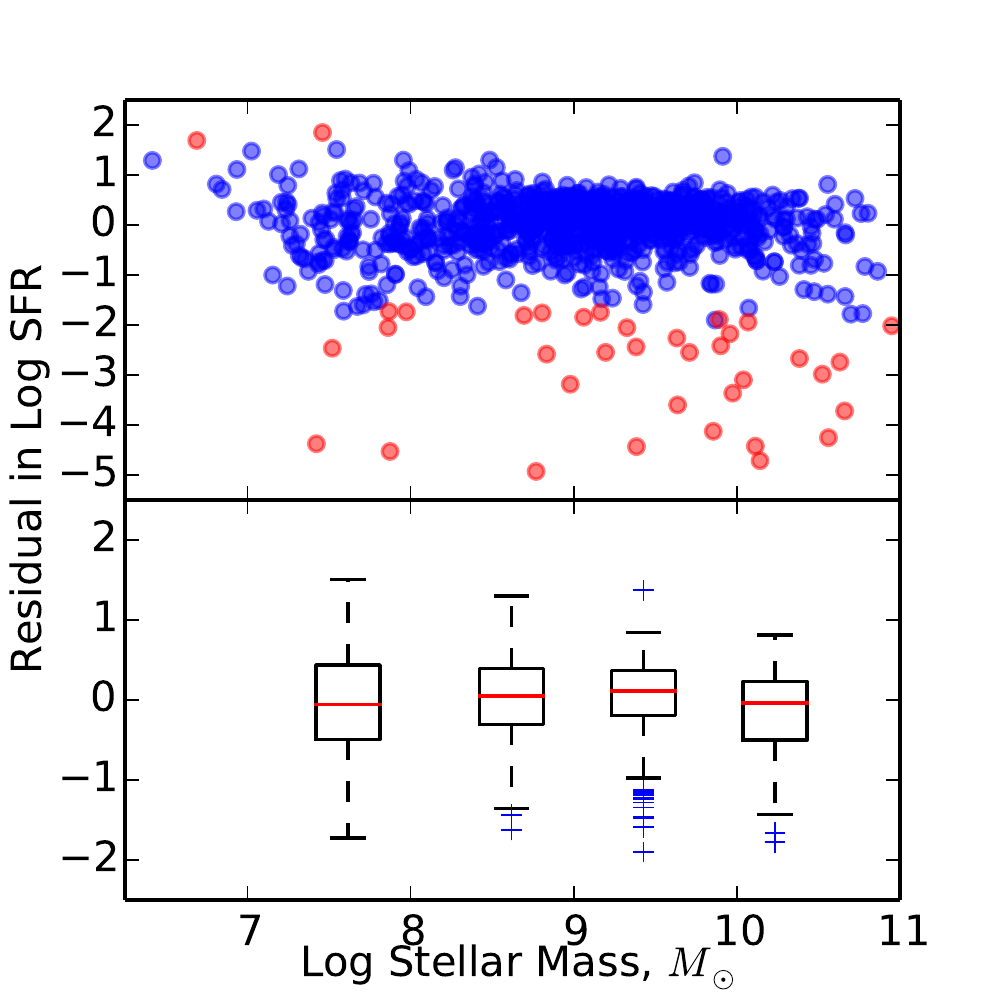}
}
\subfigure[$1.0<z\leq1.5$]{%
	\includegraphics[scale=0.40]{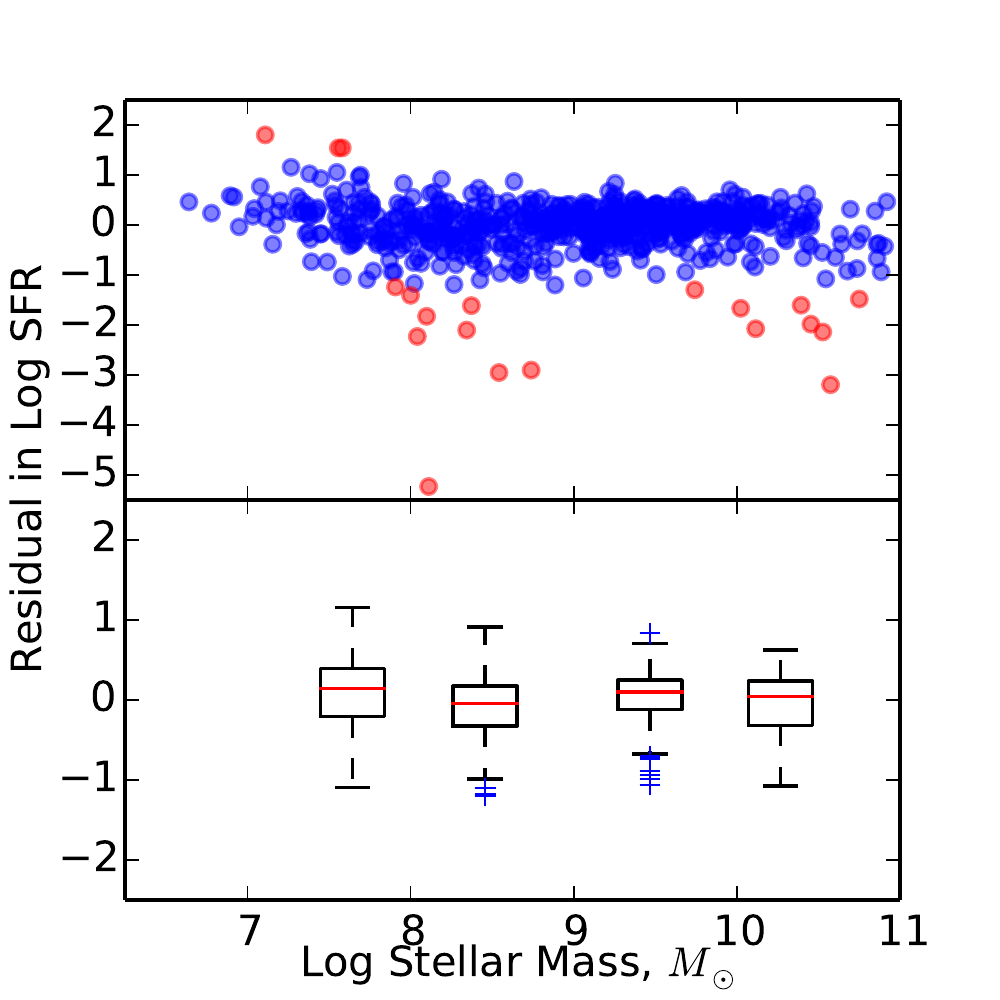}
}
\subfigure[$1.5<z\leq2.0$]{%
	\includegraphics[scale=0.40]{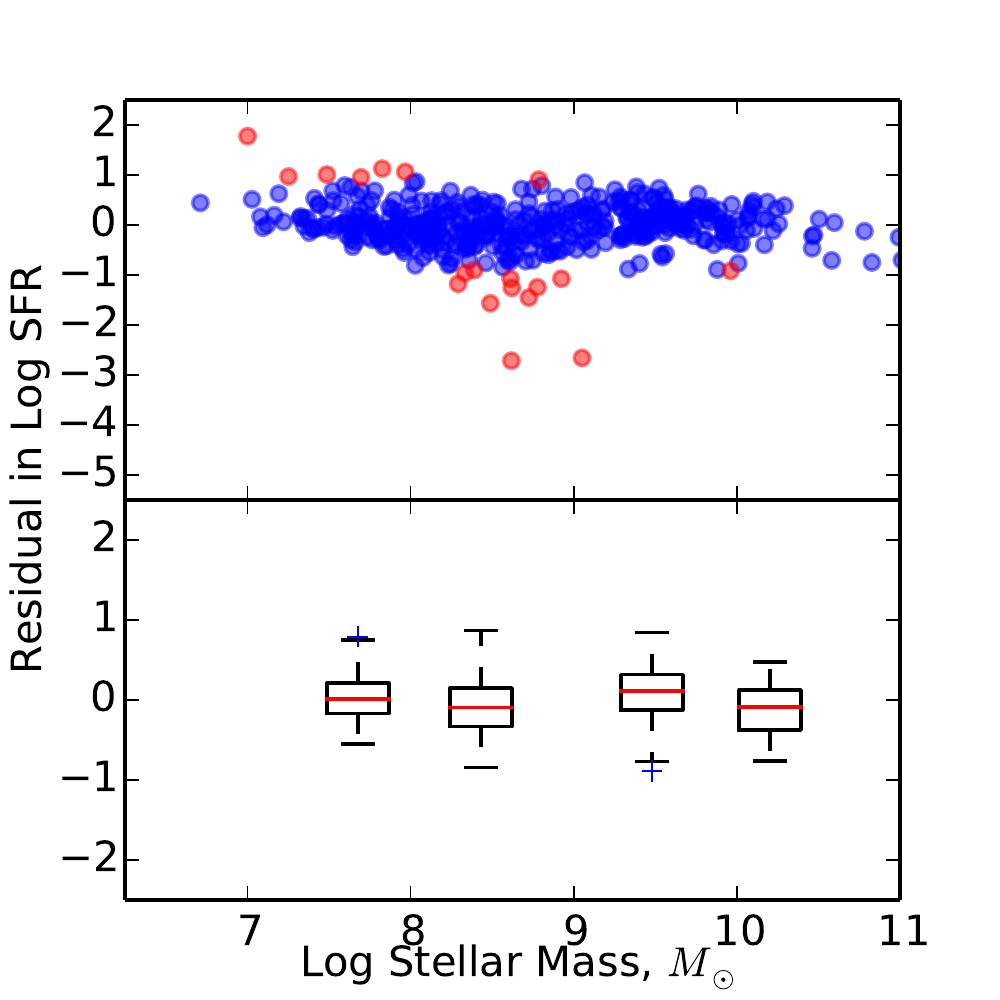}
}
\subfigure[$2.0<z\leq2.5$]{%
	\includegraphics[scale=0.40]{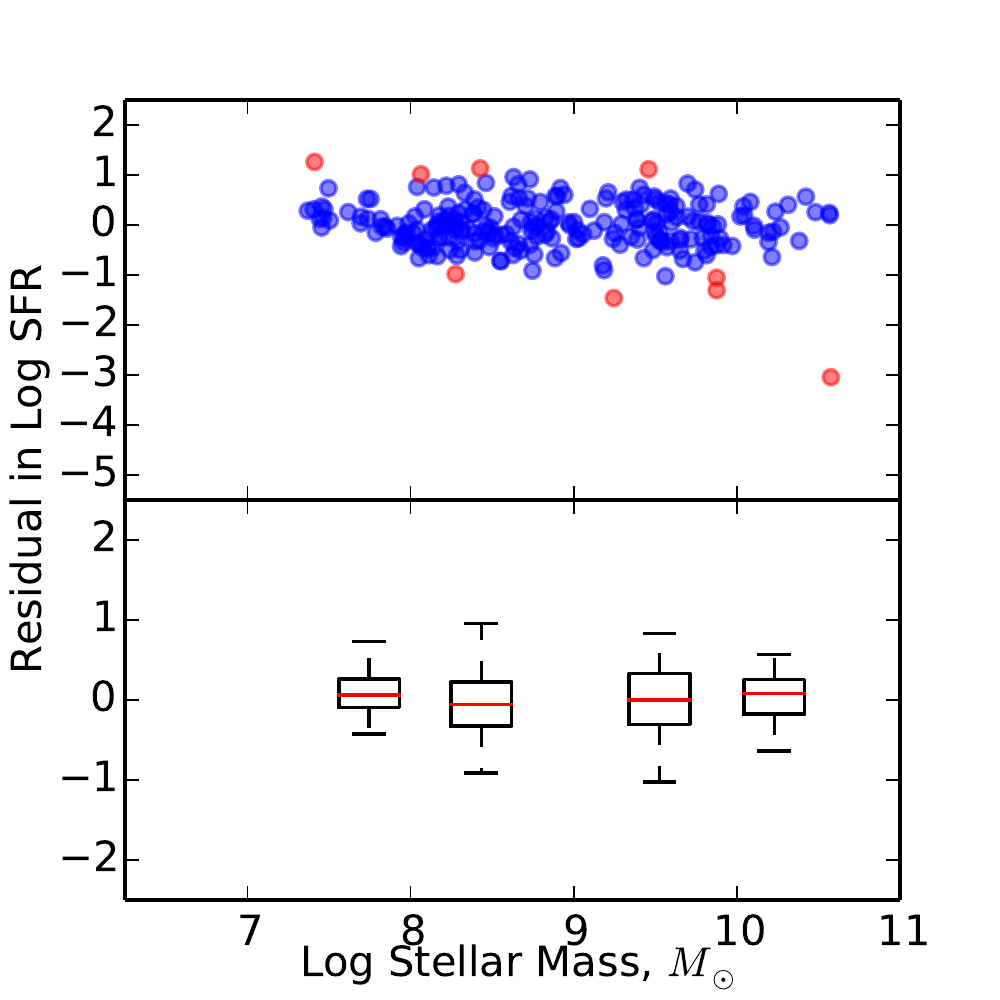}
}
\subfigure[$2.5<z\leq3.0$]{%
	\includegraphics[scale=0.40]{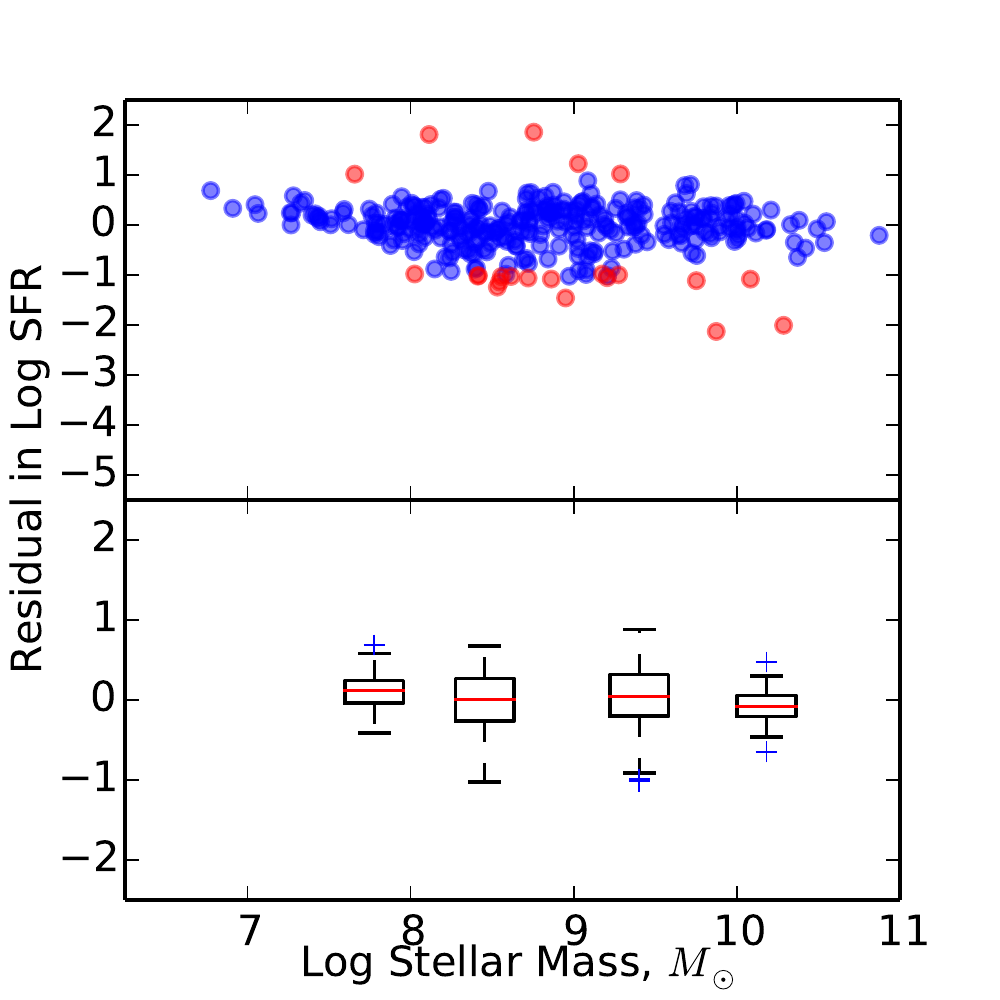}
}

\end{center}
\caption{Residuals to fits of Star Formation Rate (SFR) vs Stellar Mass (M*) shown in Figure \ref{FIGURE:01}.  {\it (Upper panels)} Residuals from initial fit are shown as blue circles, with outliers in red.  {\it (Lower panels)} Residuals are analyzed in four bins of stellar mass ($<10^8, 10^8-10^9, 10^9-10^{10}, >10^{10}$ M$_\odot$).  Box heights and whiskers indicate inter-quartile ranges (IQR) and 1.5 $\times$ IQR of the residuals in each bin.  Median residual is indicated by the red lines within each box, and horizontal box placement is at the median stellar mass of each bin.}  
\label{FIGURE:02}
\end{figure}

Figure \ref{FIGURE:02} shows that residuals do not suggest deficiencies in the model or the fits:  the band of residuals clusters around zero (suggestive of a good fit) and does not curve with M$_*$ (higher order model is not needed).  We find more negative residuals than positive residuals due to an age-gradient effect:  age decreases toward the upper left in Figure \ref{FIGURE:01}, roughly perpendicular to the best-fit line.  Consequently, there is a sharp upper cutoff in the locus of galaxies as age diminishes toward zero; older galaxies are found below and to the right.

The distributions of total scatter of the mass-binned residuals are indicated in the bottom panels of Figure \ref{FIGURE:02}.   Box plots indicate the inter-quartile ranges, and red lines indicate the medians, which are near zero.  

\begin{figure}[h]
\begin{center}
\includegraphics[scale=0.60]{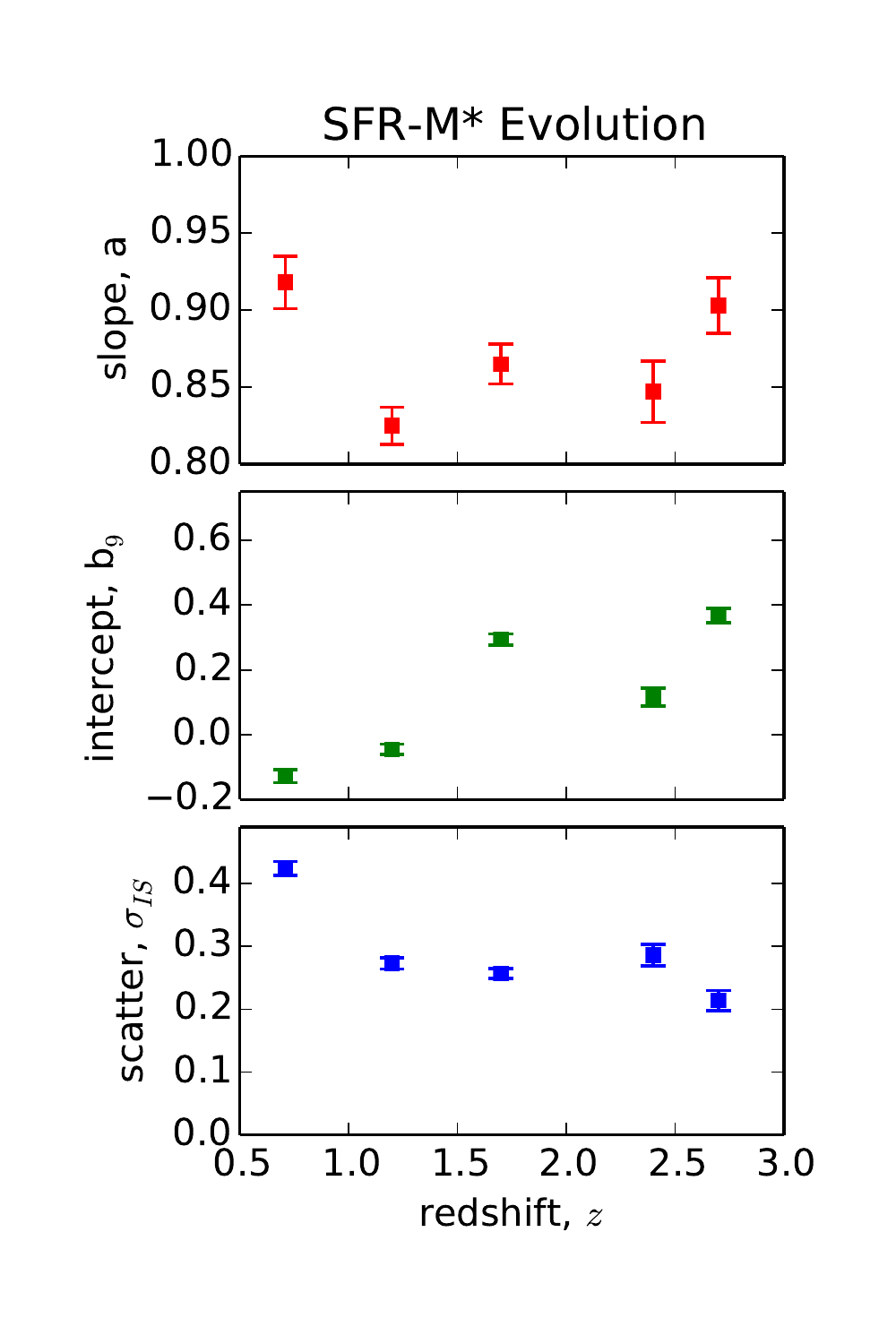}
\end{center}
\caption{Estimated model parameters for the log SFR - log M$_*$ relationship analyzed in five redshift bins in the range $0.5 < z \leq 3.0$.  Slope (top panel) and intercept (middle panel) refer to the linear components of the model, with M$_*$ values scaled to 10$^9$ M$_\odot$.  The width, $\sigma$, of the Gaussian intrinsic scatter is shown in the bottom panel, in units of dex.  Errors to the model parameters are computed from simulations.}  
\label{FIGURE:03}
\end{figure}

The estimated parameters are shown in Figure \ref{FIGURE:03} and Table \ref{TABLE:01}.  We detect intrinsic scatter in all redshift bins; scatter increases with cosmic time from the highest redshift bin to the lowest bin, from $0.220$ dex to $0.427$ dex for intrinsic scatter and $0.369$ dex to $0.525$ dex for total scatter, respectively.  The estimated slope is near unity, and we find the intercept to decrease with cosmic time, similar to trends found in \citet{2014ApJ...795..104W}. 
 
We do not find the turnover in slope above log M$_* \sim 10$ M$_\odot$ that has been previously reported \citep{2015ApJ...801...80L}; our study, aimed at low mass, has small number statistics above log M$_* > 10.5$.  Below log M$_* \sim$ 8.0, we continue to find a linear trend.

Table \ref{TABLE:02} shows intrinsic and total scatter in mass-binned sub-samples. For each sub-sample, the linear model parameters are pinned and only the intrinsic scatter is estimated.  At each redshift, the total scatter is relatively constant across the mass range; it is smallest at low mass, and relatively constant or somewhat increasing toward higher mass.  The scatter does not increase in the lowest mass bin, which is particularly surprising because, as mentioned above, scatter in SFR-M$_*$ is greater at low mass in local dwarf galaxies, and also in simulations.  Because scatter is associated with bursty star formation, these results suggest that at log M$_* \sim 7$, we do not see a significant increase in burstiness compared to higher masses.
\section{TREATMENT OF UNCERTAINTIES}
\label{ErrorsSection}

%
%
Our analysis incorporates covariances between SED fit parameters, which are non-negligible.  Median, correlated SFR and M$_*$ uncertainties are indicated as error ellipses in Figure \ref{FIGURE:01}.  M$_*$ uncertainties increase toward lower mass; SFR uncertainties and covariances exhibit no trend with mass.  SFR tends to be anti-correlated with M$_*$, \eg the $1.0 < z \leq 1.5$ SFR-M$_*$ correlation has mean $=-0.46$.  Neglecting covariances over-estimates intrinsic scatter by $\sim5-10$\%, whereas slope and intercept estimates are not significantly affected.

Uncertainties to SFR- M$_*$ model parameters are determined by simulation.  Random realizations are formed from the best-fit model; additional Gaussian random noise and intrinsic scatter are added.  Simulations have 1000 realizations, and use the same analysis as on the observed data.  Uncertainties are given by the standard deviations of the resulting true error distributions.  

%
%
%
To assess systematics, we use several fitting methods.  We use Ordinary Least Squares (OLS), Weighted Least Squares ($\chi^2$ minimization) and Orthogonal Distance Regression (ODR), although they do not use the fully available uncertainties and covariances or estimate intrinsic scatter. We also implement methods that estimate intrinsic scatter \citep{2002ApJ...574..740T,2007ApJ...665.1489K} or account for it \citep{1996ApJ...470..706A}.

We separately compute the intrinsic scatter variance, $\sigma^2_{IS}$,  from the fit residuals, ie. total scatter, $\langle \sigma^2_T \rangle$, (where $\langle \rangle$ denotes the sample mean) and the log M$_*$ and log SFR errors, $\sigma_X, \sigma_Y$ respectively, the covariances, $\textnormal{Cov}(X,Y)$, and slope, $a$, which are related as:
\begin{equation}
\sigma^2_{IS} = \langle \sigma^2_T \rangle - \langle \sigma^2_Y \rangle - a^2 \langle \sigma^2_X \rangle + 2 a \langle \textnormal{Cov}(X,Y) \rangle
\label{EQUATION:scatter}
\end{equation}
We implement this computation for methods that do not explicitly model scatter.  For large scatter, we find excellent agreement between methods (e.g. less than 2\% variation for $\sigma_{IS} \sim 0.24$) and at low intrinsic scatter we find significant dispersion (e.g. 66\% variation for $\sigma_{IS} \sim 0.08$).

To determine the systematic effect of spectroscopic redshift selection, we analyze an independent sample of number-matched photometric redshift sources from CANDELS (combined with the UVUDF sources) in the range $1.0 < z \leq 1.5$.  We find that total scatter and outlier fractions are unchanged.  However, intrinsic scatter is reduced in the photometric sample by $\sim35$\% in methods without covariances,  and $\sim 60$\% in methods that use covariances.    The total scatter is unchanged in the photometric sample, whereas the less-accurate photometric redshifts increase the scatter due to M$_*$.   Thus reduced intrinsic scatter in the photometric sample follows from Equation \ref{EQUATION:scatter}; the remaining available variance in the ``scatter budget'' available to intrinsic scatter is reduced.  This observations affirms our using the best available photometric redshifts.

We investigate whether our results may be biased by incompleteness.  We pay particular attention to low-mass galaxies (log M$_* \lesssim 9$) that are detected predominantly in UVUDF, and for which mass incompleteness sets in at $z>1$ in CANDELS data.  The UVUDF detection image is an average of eight wavebands from $F435W$ redward to $F160W$, and therefore has a complex selection function.  To approximate this function in the SFR-M$_*$ plane, we use a UVUDF flux density threshold corresponding to magnitude 29.0 in the detection image.  We use SED model parameters to express this detection threshold in terms of SFR and  M$_*$; these selection functions are shown as black curves in Figure \ref{FIGURE:01}.  We are insensitive to galaxies below and to the left of these curves.  We cannot rule out the possibility of extremely passive galaxies far from the SFR-M$_*$ correlation from having been missed; however, such galaxies would be excluded from our analysis as outliers.  Thus our results are robust to this incompleteness.  However, above $z > 2$, these curves suggest that scatter estimates at low mass are significantly affected by incompleteness.

We also investigate the dependence of our results upon the assumed form of the star formation history.  We completed analyses with SED fit parameters obtained from constant and exponential star formation histories in addition to the linear-exponential model.  For example, at $1.0 < z \leq 1.5$, intrinsic scatter is 0.13, 0.20, 0.28 dex for the constant, linear-exponential and exponential star formation histories, respectively.  Thus varying star formation history reveals a systematic uncertainty of $\pm0.08$, with constant star formation history leading to the lower value and the exponentially declining leading to the higher value; the true systematic uncertainty may be less given the unphysical assumptions of the alternate star formation histories.  

As with any parameter estimation, the fidelity of our results depends upon the efficacy of the model, which in the present case includes the assumed form of the star formation history.  We adopt the linear-exponential model because of its flexibility and good SED fits compared to available alternatives.  A logical extension of this work would be to include more complex star formation histories that include multiple bursts (our preferred, linear-exponential model effectively accommodates a single, initial burst) and determine from simulation the extent to which the data can discriminate between alternatives.

\section{CONCLUSION}
\label{ConclusionSection}

%
These results extend the study of the SFR-M$_*$ relationship of star-forming galaxies in the redshift range $0.5 < z \leq 3$ by more than an order of magnitude in stellar mass.  This lower mass limit of $\sim$10$^7$ M$_\odot$ is comparable to dwarf galaxies in the local universe.  We use SED fitting to estimate SFR and M$_*$ as well as their uncertainties and covariances.  Where measurable, we find the intrinsic scatter to be a substantial fraction ($\gtrsim$ 50\%) of the total scatter.  We find the intrinsic scatter to be $\sigma_{IS} \approx0.2-0.4$ dex, see Tables \ref{TABLE:01} and \ref{TABLE:02}.  These values are somewhat larger than the simulations of \citet{2010MNRAS.405.1690D}, who find $\sigma = 0.11$ dex at $z\sim0$, but are in good agreement with the Illustris simulations in the overlapping mass range M$_* > 10^{8}$ M$_\odot$  below $z \leq2$ \citep{2015MNRAS.447.3548S}.  We encourage modelers to report their observables to even lower mass for comparison with these observations.

We find the intrinsic scatter in the SFR-M$_*$ relation to increase with cosmic time (decreasing redshift) by about a factor of two across the range $2.5 > z > 0.5$, although most of this increase occurs for a single redshift bin, $0.5 < z \leq 1.0$.  Increasing scatter with cosmic time is also found in the models of \citet{2015MNRAS.453.4337S} and \citet{2015MNRAS.447.3548S}.

At each redshift, we find the scatter to be relatively constant (or slightly decreasing) toward lower mass, particularly above $z>1$, in disagreement with trends for broadband SFR reported in the theoretical studies of \citet{2015MNRAS.451..839D} and \citet{2015arXiv151003869S}.  These studies each report substantially larger scatter at low masses for H$\alpha$ based SFRs than the broadband ones used here.  SED fitting is sensitive to $\gtrsim$100~Myr timescale variability, while spectroscopic indicators are needed for shorter time variability.  We interpret the absence of increased scatter to mean that such intermediate or long timescale variability does not dominate the star formation histories of low-mass galaxies. 

Without specifically addressing the timescale issue, the simulations of \citet{2015MNRAS.453.4337S}, predict a moderate increase in scatter toward low mass in the range M$_* > 10^8$ M$_\odot$ over our redshift range, whereas the simulations of \citet{2015MNRAS.447.3548S} and \citet{2010MNRAS.405.1690D} show constant scatter with mass down to M$_* = 10^9$ and M$_* = 10^8$ M$_\odot$ respectively.  In these simulations, SFR is computed from molecular hydrogen gas density and empirically motivated models of sub-grid physics.

The origin of the confinement of star-forming galaxies to a narrow SFR-M$_*$ correlation is a theoretical question of major interest (\eg \citealt{2010MNRAS.405.1690D,2015arXiv150902529T,2015arXiv150804842R} and references therein).  \citet{2015arXiv150902529T} show that it could be understood in terms of the evolution of galaxies through phases of gas compaction, depletion, possible replenishment, and eventual quenching.  In any case, the low scatter we observe suggests a remarkable consistency in star formation spanning 3-4 orders of magnitude in galaxy stellar mass.  It invites comparison with other dynamical systems across a variety of disciplines from physics to biology where power law scaling relations are associated with self-regulating dynamics.
 
\acknowledgments

The authors wish to thank Aaron Clauset for discussions.  Support for HST Program GO-12534 was provided by NASA through grants from the Space Telescope Science Institute.  This material is based upon work supported by the National Science Foundation under grant no. 1055919.  This work is based on observations taken by the CANDELS Multi-Cycle Treasury Program with the NASA/ESA HST, which is operated by the Association of Universities for Research in Astronomy, Inc., under NASA contract NAS5-26555. 

%
%
\bibliographystyle{apj}                       
\bibliography{apj-jour,sfr_scatter_evolution}
%
%

%
%
%
%
%
%


\clearpage

%
%
%
%
%
%


%
%
%
%


%
%
%
\begin{deluxetable}{lcccccccc} 
\tablecolumns{9} 
\tablewidth{0pc} 
\tablecaption{Linear Plus Intrinsic Scatter Model Parameters} 
\tablehead{ 
\colhead{Redshift} & \colhead{$N$} & \colhead{$a$} &  \colhead{$b$} &\colhead{$b_{9}$} & \colhead{$\sigma_{int}$} &\colhead{$\sigma_{Tot}$}\\ 
\colhead{(1)} & \colhead{(2)} & \colhead{(3)} & \colhead{(4)} & \colhead{(5)} & \colhead{(6)} & \colhead{(7)} }
\startdata
$0.5<z\leq1.0$	&	913	&	0.919$\pm$0.017	&	-8.394$\pm$0.011	&	-0.121$\pm$0.021	&	0.427$\pm$0.011	&	0.525\\
$1.0<z\leq1.5$	&	671	&	0.825$\pm$0.012	&	-7.474$\pm$0.010	&	-0.045$\pm$0.016	&	0.273$\pm$0.009	&	0.383\\
$1.5<z\leq2.0$	&	447	&	0.867$\pm$0.013	&	-7.484$\pm$0.011	&	0.321$\pm$0.017	&	0.255$\pm$0.008	&	0.354\\
$2.0<z\leq2.5$	&	237	&	0.849$\pm$0.021	&	-7.513$\pm$0.018	&	0.128$\pm$0.028	&	0.281$\pm$0.017	&	0.399\\
$2.5<z\leq3.0$	&	304	&	0.899$\pm$0.017	&	-7.729$\pm$0.015	&	0.367$\pm$0.023	&	0.220$\pm$0.017	&	0.369\\
\enddata

\tablecomments{(1) Redshift range of the sample.  (2) Number of galaxies in the final fit (excluding outliers).  (3,4,6) Estimated parameters of the model $\textnormal{log}~SFR = a~\textnormal{log}M_* + b + N(0,\sigma_{int})$ including SFR and M$_*$ uncertainties and covariances.  (5) Intercept, b$_9$, corresponds to the mass-scaled model $\textnormal{log}~SFR = a (\textnormal{log} M_*- 9.0) + b_{9}$ in which errors to the fit parameters are approximately uncorrelated. (7) Total scatter, defined as sample standard deviation of the fit residuals after clipping of outliers.}
\label{TABLE:01}
\end{deluxetable} 

%
%
%
\begin{deluxetable}{lcccc} 
\tablecolumns{5} 
\tablewidth{0pc} 
\tablecaption{Scatter about the SFR-M$_*$ relation for galaxies in CANDELS/UVUDF in bins of stellar mass and redshift.} 
\tablehead{ 
\colhead{Statistic} & \colhead{$6 < \textnormal{log}M_* \leq 8$} & \colhead{$8 <\textnormal{log}M_* \leq 9$} & \colhead{$9 < \textnormal{log}M_* \leq10$} & \colhead{$10 < \textnormal{log}M_* \leq 11$}}
\startdata
\multicolumn{5}{c}{$0.5 < z \leq 1.0$}  \\
\hline \\
Num Galaxies	   & 128		&	298	&		430	&		102\\
Intr.~scat., dex & 0.462$\pm$0.030	&	0.404$\pm$0.012	&	0.315$\pm$0.011	&	0.435$\pm$0.026\\
Total scat., dex  & 0.552	&		0.445	&	0.368	&	0.428\\
Outlier fraction $+$ & 0.031	&		0.000	&	0.000	&	0.000\\
Outlier fraction $-$ & 0.156	&		0.067	&	0.074	&	0.176\\
\hline \\
\multicolumn{5}{c}{$1.0 < z \leq 1.5$}  \\
\hline \\
Num.~galaxies	   & 111	&		209	&		284	&		87\\
Intr.~scat., dex  & 0.201$\pm$0.025	&	0.249$\pm$0.010	&	0.230$\pm$0.006	&	0.281$\pm$0.014\\
Total scat., dex  & 0.315	&		0.285	&	0.285	&	0.348\\
Outlier fraction $+$& 0.027	&		0.000	&	0.000	&	0.000\\
Outlier fraction $-$ & 0.009	&		0.053	&	0.004	&	0.103\\
\hline \\
\multicolumn{5}{c}{$1.5 < z \leq 2.0$}  \\
\hline \\
Num.~galaxies	   & 99	&		189	&		144	&		30\\
Intr.~scat., dex & 0.279$\pm$0.022	&	0.497$\pm$0.018	&	0.332$\pm$0.008	&	0.417$\pm$0.025\\
Total scat., dex  & 0.406	&		0.437	&	0.340	&	0.348\\
Outlier fraction $+$ & 0.000	&		0.000	&	0.000	&	0.000\\
Outlier fraction $-$ & 0.000	&		0.005	&	0.007	&	0.000\\
\hline \\
\multicolumn{5}{c}{$2.0 < z \leq 2.5$}  \\
\hline \\
Num.~galaxies	   & 29	&			111	&		85	&		18\\
Intr.~scat., dex  & 0.232$\pm$0.050	&	0.337$\pm$0.027	&	0.425$\pm$0.022	&	0.240$\pm$0.048\\
Total scat., dex  & 0.354	&		0.451	&	0.491	&	0.308\\
Outlier fraction $+$& 0.000	&		0.000	&	0.000	&	0.000\\
Outlier fraction $-$ & 0.000	&		0.000	&	0.000	&	0.053\\
\hline \\
\multicolumn{5}{c}{$2.5 < z \leq 3.0$}  \\
\hline \\
Num.~galaxies	   & 50	&			146	&		106	&		23\\
Intr.~scat., dex  & $<0.03 (3\sigma)$	&	0.421$\pm$0.018	&	0.392$\pm$0.019	&	0.309$\pm$0.069\\
Total scat., dex  & 0.267	&		0.516	&	0.464	&	0.331\\
Outlier fraction $+$& 0.000	&		0.000	&	0.000	&	0.000\\
Outlier fraction $-$ & 0.000	&		0.000	&	0.009	&	0.042\\
\enddata

\tablecomments{The number of galaxies in each bin is tabulated, and the intrinsic scatter is estimated with the method of F87 with errors determined from simulation.  The total scatter (standard deviation of residuals to the linear fit) is tabulated for comparison.  Outlier fraction $+$ ($-$) refers to the fraction of sources in each bin above (below) the initial best-fit line that are clipped and excluded from the final fit.}
\label{TABLE:02}
\end{deluxetable} 

\end{document}